\magnification=\magstep1
\baselineskip=14truept
\font\tif=cmr10 scaled \magstep3

\rightline{PUPT-1596$|$cond-mat/9602123}
\vfil
\centerline{\tif Critical theory of turbulence in 
incompressible fluids}
\vfil
\centerline{{\rm Vipul 
Periwal}\footnote{${}^\dagger$}{vipul@puhep1.princeton.edu}}
\bigskip
\centerline{Department of Physics}\centerline{Princeton University
}\centerline{Princeton, New Jersey 08544-0708}
\vfil
\par\noindent The Navier-Stokes equation describes the deterministic
evolution of incompressible fluids.  The effects of random initial
conditions on solutions of this equation are studied.  It is shown 
that there is an infrared stable
fixed point accessible within the epsilon expansion, about $d_c=4,$
with a particular choice of viscosity.  The behaviour of the usual
viscous fluid is obtained as a relevant perturbation about this 
fixed point, with the usual viscosity serving as $T-T_c.$  
\medskip
\centerline{PACS: 47.25-c, 11.10}
\footline={\hfil}
\vfil\eject
\footline={\hss\tenrm\folio\hss}

\def\part{\partial}
\def\wl{2}
\def\t{1}
\def\jc{3}
\def\fluid{4}
\def\zj{5}
\def\ll{6}

\def\partt{\partial_t}
\def\vu{\vec u}
\def\Del{\Delta}
The equations of fluid mechanics are deterministic.  
The motion of an incompressible fluid, in the absence of viscous forces, is 
governed by
$$ \partt \vu + \vu\cdot \nabla \vu = - {1\over\rho}\nabla p,\eqno(1)$$
and the incompressibility condition, $\nabla\cdot\vu=0.$  
Here $\vu$ is the velocity of a fluid element, $p$ is the pressure,
and $\rho$ is the constant density of the fluid.
In fact, the pressure term is determined up to a harmonic function by
the velocity field using $\nabla\cdot\vu=0.$
This equation has no obvious expansion parameter, since all the terms,
with precisely determined coefficients, are determined by Newton's
equation, $F=ma$---as a corollary, the equation possesses Galilean invariance.

In principle,
all details of the fluid's motion are determined by completely defined
initial conditions.  The ideal aim of a theory of turbulent motion
would be to establish a statistical description analogous to the kinetic theory
of gases, based on the deterministic dynamics of fluids.  The statistical
description may  be thought of as an ensemble average[\t], over different
experiments corresponding to different initial conditions.  
It is the aim of this paper to perform such an average over random initial
conditions for incompressible fluids.

The approach taken here is analogous to the standard infrared stable 
non-trivial fixed 
point found in the $\epsilon$ expansion for $\phi^4$ theory, 
about $d_c=4$[\zj].
In that case, one considers the massless theory about $d_c=4,$ expands
in $\epsilon=d_c-d,$ and one finds a fixed point with $\epsilon\propto g.$
Correlation functions at $T>T_c,$ in the critical domain, {\it i.e.} with
the correlation length large compared to the cutoff, but still finite,
exhibit strong scaling. Physics at length scales small compared to
the correlation length, but large compared to the underlying cutoff,
can be expressed in terms of
universal correlation functions in which the unit of length is 
provided by the correlation length.  

Turbulence, in 
general, involves the transfer of energy from large-scale mean motions 
to viscous dissipation at small scales.  The case of {\it 
isotropic homogeneous freely decaying 
turbulence} considered in this paper, 
however, involves no energy production[\t], and hence one need
not take into account the so-called `energy cascade', only
viscous dissipation at small scales.    
Ultimately, all
turbulence dies away due to non-universal dissipative viscous forces, 
so the theory of interest is only expected to
apply in the inertial regime where the viscous forces dominant at
long distances can be neglected.  
These dominant viscous forces  
determine the length scale of decay, so in the inertial
range we are interested in fluid velocity correlations at distances
long compared to the underlying molecular cutoff, but short compared
to this  viscous length scale.  Compare this description
of the inertial range with the description given above of $\phi^4$ theory
with $T-T_c$ small but non-zero.  In the inertial range then, the viscous
forces of interest are {\it higher} terms in the expansion of viscous forces in
powers of wavenumbers (and, in principle, nonlinear terms in the velocity 
itself---these will turn out to be irrelevant).  We will
find that there is a critical point in this `hyperviscous' theory, with 
the viscous theory {\it in the inertial range} described by 
universal correlation functions in which the unit of length is
provided by the standard viscosity.  In this sense, our approach is
modelled closely on the successful $\epsilon$ expansion framework of
critical phenomena.  

Now, while the operational meaning of the ensemble average of
physical interest was described above, previous 
theoretical investigations in a field-theoretic framework 
have added a `random force'
term to the basic equation---the resulting exponents of turbulent flow
depend strongly on the scaling form assumed for the force-force correlation
function, contradicting the observed universality of the exponents[\fluid].
By the fluctuation dissipation theorem, the amplitude of the 
random force has to be related to the viscous forces for thermal equilibrium 
to be reached at long time scales.  This correlation implies that the 
dimensionality of the order parameter (in this case the velocity field)
depends on the random force.  This is an explanation for the 
non-universality of theoretical computations[\wl].

Several years ago, Liao[\wl] examined the effects of averaging over all
solutions of eq.~(1) in a field theoretic setting, and obtained
significant results on the spectra of fully developed isotropic turbulence.
My analysis is similar in spirit to Liao's approach, but is based on 
Cardy's calculational framework for the effects of random initial
conditions on deterministic diffusion equations with nonlinear terms[\jc].

The initial conditions I consider are the simplest allowed by
locality and incompressibility:
$$\langle u^i(x,0)u^j(y,0) \rangle = - g (\Del\delta^{ij} - \part^i\part^j)
\delta(x-y),$$  
with $\Del\equiv \nabla^2.$  It is important to keep in mind that there is
an underlying cutoff in the problem, even though our computations will
be done with dimensional regularization.  
I shall assume $\langle u^i(x,0)\rangle =0.$
From these initial conditions, and eq.~(1), observe the 
following engineering dimensions: 
$[u] = Ek^{-1}, [g] = [u]^2 k^{-(d+2)}.$
In an isotropic incompressible 
homogeneous fluid, the viscous forces are the only non-universal terms
in the equation.  I am interested in the inertial regime, where
the viscous forces are negligible, yet the form of the viscous terms
determines the dimension of $[u].$  Assume that the viscous force is of the
form $\nu_\beta \Del^\beta \vu + {\rm lower\ order},$ 
Then $[\nu_\beta]
=Ek^{-2\beta}.$  In principle, one could eschew the analyticity in momenta 
assumed implicitly here, but the framework of renormalizability must then
be considered {\it ab initio}.  
Following Cardy[\jc], I expect that the only quantity
that requires renormalization will be $g,$ since the evolution equation
is deterministic, and the only functional average performed is 
over the initial conditions.  I will henceforth set $\nu_\beta=1.$
Then
$$[g] = k^{4(\beta-1)-d},$$
so for $d_c=4,$ $\beta=2.$  
As explained above, this is what we expect in order to find the critical
point suited to a description of the inertial regime: 
The standard viscous force, $\nu \Del \vu,$ will appear in our theory
as a `massive' deformation away from the critical theory, with the 
hyperviscous force, $\nu_2 \Del^2 \vu,$ determining the critical theory.
Of course, a standard mass term would violate Galilean invariance
for a fluid, so in fact this analogy is quite precise.

As mentioned above, 
the compatibility of  eq.~(1) with the incompressibility
conditions requires that $p$ is related
to $\vu$ by
$$ - {1\over\rho}\Del p = \part_iu_j\part_ju_i,$$
in order that initial conditions satisfying $\nabla\cdot\vu(x,t=0)=0$
produce a flow satisfying $\nabla\cdot\vu(x,t)=0.$  This expression is
dimensionally correct[\t]. (Compare with Landau and 
Lifshitz[\ll, eq.~(31.4)].) 
Only the gradient of the pressure enters eq.~(1),
so there is no ambiguity in inverting the Laplacian for our purposes.
The initial conditions are explicitly local, the inversion of the
Laplacian needed for the pressure term is merely a consequence of the
infinite velocity of sound in an incompressible fluid.

Treat the nonlinear terms in eq.~(1) formally as perturbations.  This 
leads to an iterative solution to the equation which has the structure 
of a tree graph, as is standard[\zj].
The trunk of the tree represents the 
solution at time $t,$ $\vu(x,t),$ and the time-ordered branches all terminate
at $\vu(x_i,0).$
The ensemble average of interest has
the effect of closing the branches into loops, with attendant factors of
$g.$  It is then clear 
that $g$ is the expansion parameter of interest---remarkably, the ensemble
average has a perturbative expansion, even though the actual equation has
no such perturbative parameter.  
\def\ee{{\rm e}}
\def\dd{{\rm d}}

Define $\Pi^{il}(k) \equiv \delta^{ij} - k^ik^j/k^2.$  The formal solution
of the equation then has the following structure:
$$u^i(k,t) = \ee^{-tk^4}u^i(k,0) - i\Pi^{il}(k)k^j\int_{0}^{t}\dd t'
\dd p \ee^{-(t-t')k^4}
\ee^{-t'(p^4+(k-p)^4)}u^l(p,0)u^j(k-p,0) + \dots.\eqno(2)$$
Any correlation function of the form $\langle \prod u^{k_i}(x_i,t_i)\rangle$
can be evaluated in a straightforward fashion, 
using the assumed form of the initial 
conditions.  To find a fixed point, the renormalization of $g$ must
be computed.  It is
elementary to observe that at order $g^p$ the degree of divergence of
an $n$ point function is 
$$\delta = d - (d-3)n + (d-4)p.$$
It would appear then that there are three possible primitively divergent
functions at $d=4$ (since the one point function is trivial by Galilean 
invariance).  However, this is not actually the case, due to the 
incompressibility condition.  This condition has a drastic effect on
the vertices, resulting in explicit factors of momentum coming out
of loop integrals, apparent in eq.~(2).
The actual degree of divergence 
of the two point function is only logarithmic, and the other functions 
are actually finite.  The divergent function can be 
rendered finite with just a renormalization of $g$, due
to this fact.  A corollary is that the case of a compressible fluid is not
a trivial extension of the present approach.
\def\eps{\epsilon}

The integrals that arise in loop calculations for arbitrary times are
unfortunately not easily performed, due to the quartic character of 
the hyperviscous force.  To compute renormalization group functions, 
I need only the explicit value at $E=0,$ going to an energy-momentum
representation.  
The divergent contribution to the two point function at $O(g^2)$ comes from
the integral
$$I\equiv \int {\dd^{d}p \over (2\pi)^d} {p^2(q-p)^2\Pi^{lr}(p)\Pi^{ms}(q-p) %
\over {\left(p^4 + (q-p)^4\right)^2}}.$$
There are other contributions at $O(g^2)$ that are manifestly finite,
and difficult to evaluate.
The two point function (excluding uninteresting
finite parts at $O(g^2)$) is 
$(\eps\equiv 4-d)$
$$\langle u^i(q,E=0)u^j(-q,E=0)\rangle\Big|_{q^2=\mu^2} 
= 16\pi^2g{\Pi^{ij}(q)\over q^{6}}
\left(1 + C{g} {\mu^{-\eps}\over \eps}\right).$$
I have rescaled $g$ to absorb factors of $16\pi^2,$ and $C=7/24.$
Define the dimensionless renormalized coupling by 
$g_r \equiv g\mu^{-\eps}\left(1 + C{g} {\mu^{-\eps}\over \eps}\right),$
then 
$$\mu{\part\over\part \mu}g_r \equiv \beta(g_r) = -\eps g_r - C{g_r^2} $$
There is a zero of the beta function at $g_r^* = - 
\eps/C,$ with $\beta'(g_r^*) = \eps>0,$ implying infrared stability.
The sign of $g_r^*$ is important: The fact that it is negative signifies 
that the critical point is associated with a multi-modal distribution,
which is physically quite appropriate for turbulence---we do not
observe turbulent behaviour starting spontaneously in beakers of water from
thermal fluctuations, an example of a {\it unimodal} set of 
initial conditions.

Various simple scaling properties at the critical point are 
easily derived in the standard renormalization group framework[\jc].
For example, at fixed $q,E,$
$$\langle u(q,E)u(-q,E)\rangle \sim  \mu^{\eps - \gamma^*}g_r^*,$$
with 
$\gamma \equiv (\beta + \eps g_r)/g_r,$ and 
$\gamma^*\equiv\gamma(g_r^*)=\eps.$ 
It follows that 
$$\langle u(x,t)u(x,t)\rangle \sim t^{-1},$$
independent of $\epsilon.$  Similarly, one deduces that
$$\langle u(x_1,t) \dots u(x_n,t)\rangle = t^{-3n/4} 
G^{(n)}(|x_i-x_j|t^{-1/4}),$$
which is essentially just dimensional analysis taking into account the 
fact that $u$ is not renormalized.  Notice that in principle we could also 
have found universal exponents by considering the response functions 
corresponding to variations of the initial conditions, as studied in
Ref.~\jc.  In the present case, the incompressibility
condition implies that these functions are finite, hence do not exhibit
any nontrivial scaling.  The study of these response functions may 
lead to a better understanding of the convergence of the $\eps$ expansion
in this theory.

The physical case of 
interest is really the behaviour of the theory away from the critical
point, perturbed by the usual viscosity term.  As is quite standard[\zj]
the  behaviour of correlation functions at distance scales
small compared to the finite correlation length, but long compared to
the cutoff, is obtained from the critical theory with appropriate resummed
operator insertions.  This analysis poses no new conceptual problems in
the present theory, but is computationally rather involved due to the
quartic energy denominators.  However, even if the integrals are analytically
intractable, the numerical solution at $\eps=1$ should be quite feasible.

In summary, I have presented here concrete
evidence within the standard $\epsilon$
expansion for a nontrivial  infrared stable 
fixed point governing, in the inertial range of interest,
statistical properties of 
solutions of incompressible fluids when the initial conditions are
multi-modal, without the introduction of any random force.
Since numerical techniques to cope with the $\epsilon$ expansion are
very well studied, it should be possible to derive detailed predictions
for experimental observations of universal behaviour in turbulent flows
starting from correlation functions at this fixed point.
These issues, and especially the extension to
compressible fluids, will be discussed elsewhere.
Numerical simulations of fluids could be directly used to test the
hyperviscous fixed point at criticality.

\bigskip
Acknowledgements:  
I am grateful to J. Cardy for
comments on the manuscript, and W. Liao for many explanations.  
I thank D. Gross,
B. Nachtergaele, A. Polyakov and W. Taylor  for helpful conversations.
This work was supported in part by NSF Grant No. PHY90-21984.
\hfuzz=2pt
\bigskip
\centerline{References}
\bigskip
\item{\t} D.J. Tritton, {\sl Physical Fluid Dynamics}, Van Nostrand Reinhold, 
Cambridge U.K. (1977)

\item{\wl} W. Liao, {\it J. Phys. A} {\bf 22}, L737-741 (1989); {\it J.
Stat. Phys.} {\bf 65}, 1 (1991)

\item{\jc} J.L. Cardy, {\it J. Phys. A} {\bf 25}, 2765 (1992)

\item{\fluid} D. Forster, D.R. Nelson and M.J. Stephen, {\it Phys. Rev.} 
{\bf A16}, 732 (1977); C. De Dominicis and P.C. Martin, 
{\it Phys. Rev.} {\bf A19}, 419 (1979);
V. Yakhot and S.A. Orszag, {\it J. Sci. Comp.} {\bf 1}, 3 (1986);
see also S.F. Edwards, {\it J. Fluid Mech.} {\bf 18}, 239 (1964). Some results
on pressure-less fluids in one dimension can be found in A.M. Polyakov,
{\sl Turbulence without pressure}, Princeton report PUPT-1546 (1995).

\item{\zj} J. Zinn-Justin, {\sl Quantum field theory and critical phenomena},
Oxford University Press, Oxford U.K. (2nd ed., 1993)

\item{\ll} L.D. Landau and E.M. Lifshitz, {\sl Fluid Mechanics}, Pergamon
Press, Oxford U.K. (1959)

\end